\documentclass[12pt,preprint]{aastex}                         
\usepackage{bbm}                                              
\usepackage{rotating}                                         
\usepackage[norefs,nocites]{refcheck}                         
\usepackage{amsmath}                                          
\usepackage{natbib}                                           
\usepackage{epsfig}                                           
\usepackage{multirow}                                         

\newcommand{\beq}{\begin{eqnarray}}                           
\newcommand{\eeq}{\end{eqnarray}}                             

\shorttitle{Dark Energy and models }        
\shortauthors{Yue-Liang Wu, Qing-Jun Zhang et al.}            

\begin{document}                                              

\title{Modelling Time-varying Dark Energy \\ with Constraints from Latest Observations}
\author{Qing-Jun Zhang$^{1,2}$ and Yue-Liang Wu$^{2}$}

\affil{ $^1$ \rm Department of Physics, Graduate University,
\\ Chinese Academy of Sciences, Beijing 100049
China}

\affil{ $^2$ \rm   Kavli Institute for Theoretical Physics China,
\\ Key Laboratory of Frontiers in  Theoretical Physics, \\ Institute of
Theoretical Physics, Chinese Academy of Sciences, Beijing 100190, China}

 \email{ylwu@itp.ac.cn}

\begin{abstract}
We introduce a set of two-parameter models for the dark energy equation of state (EOS) $w(z)$ to investigate
time-varying dark energy. The models are classified into two types according to their boundary behaviors at
the redshift $z=(0,\infty)$ and their local extremum properties. A joint analysis based on four observations
(SNe + BAO + CMB + $H_0$) is carried out to constrain all the models. It is shown that all models get almost
the same $\chi^2_{min}\simeq 469$ and the cosmological parameters $(\Omega_M, h, \Omega_bh^2)$ with the
best-fit results $(0.28, 0.70, 2.24)$, although the constraint results on two parameters $(w_0, w_1)$ and
the allowed regions for the EOS $w(z)$ are sensitive to different models and a given extra model parameter.
For three of Type I models which have similar functional behaviors with the so-called CPL model, the
constrained two parameters $w_0$ and $w_1$ have negative correlation and are compatible with the ones
in CPL model, and the allowed regions of $w(z)$ get a narrow node at $z\sim 0.2$. The best-fit results
from the most stringent constraints in Model Ia give $(w_0,w_1) = (-0.96^{+0.26}_{-0.21}, -0.12^{+0.61}_{-0.89})$
which may compare with the best-fit results $(w_0,w_1) = (-0.97^{+0.22}_{-0.18}, -0.15^{+0.85}_{-1.33})$
in the CPL model. For four of Type II models which have logarithmic function forms and an extremum point,
the allowed regions of $w(z)$ are found to be sensitive to different models and a given extra parameter.
It is interesting to obtain two models in which two parameters $w_0$ and $w_1$ are strongly correlative and
appropriately reduced to one parameter by a linear relation $w_1 \propto (1+w_0)$.
\end{abstract}

\section{INTRODUCTION}

The distance determinations at cosmological scales, from standard candles
such as Type Ia supernovae(SNe) (Perlmutter et al. 1999, Riess et al.
1998) as well as standard rulers such
as the cosmic microwave background(CMB)(Hinshaw et al. 2009; Komatsu et al. 2009)
and  the baryon acoustic oscillation(BAO) (Eisenstein et al. 2005; Cole et al. 2005;
 Huetsi 2006; Percival et al. 2007), have firmly established that the expansion
  of our universe in the time close
to the present was accelerating.  It  is now believed that a  mysterious
negative pressure component in the universe, named dark energy, is responsible
for this accelerated expansion, which contributes about $70\%$ critical density
of universe today. The underlying physics of dark energy remains obscure and
most of relevant studies parameterize dark energy by its equation of state (EOS)
parameter $w = p/\rho$. In observational data analysis, it is widely
assumed  that the dark energy  EOS is independent on specific theoretical
models and follows a certain  predetermined evolutionary history. Common
parameterizations include the constant $w$, the linear variation of the redshift
$w(z)=w_0+w_z  z$, a convergent form $w(a)=w_0 + w_a (1-a)$ with $a= 1/(1+z)$ the scale
factor, and the piecewise functions binned on the redshift. Despite the fact that
the accumulated observational data have significantly improved the constraint on
the dark energy EOS, our knowledge about the time evolution EOS $w(z)$ is rather poor
in considering its infinite dimensional parameter space. Thus, it is useful to study
more general classes of models with the time-varying dark energy. Here we shall
utilize the 397 SNe Ia provided by Hicken et al. (2009), 7-year WMAP observations,
the BAO data given by SDSS DR7, together with the Hubble constant $H_0$ (Riess et al. 2009)
constrained from the SHOES team to investigate varieties of the parameterizations of dark energy EOS
in the flat cosmology.

By far, all observed data are consistent with the $\Lambda {\rm
CDM}$ cosmology, with dark energy in the form of  Einstein's cosmological
constant $\Lambda$ or the vacuum energy, which is time independent and has
an EOS $w\equiv -1$. However, this model suffers from the difficulties related to the
 fine tuning and the coincidence problem (see e.g. Padmanabhan 2003). In last decades,
several kinds of dynamical dark energy models have been proposed, such as
holographic, quintessence, phantoms, K-essence, branes, Chaplygin gas, etc.
In these situations, dark energy EOS changes with time and is generally parametrized
as $w(z)=w_0 + w_1 \; f(z)$, with $w_0$ the present value of dark energy EOS for f(0) = 0.
Conveniently, it includes the case of the $\Lambda {\rm CDM}$ model ($w_0 = -1,w_1 = 0$),
and the constant EOS ($w_0 = w,w_1 = 0$).

The constant EOS is the simplest choice of dark energy parameterizations
and a well approximation of the slowly rolling quintessence field (Huey et al. 1999;
Wang et al. 2000). When determining whether or not the dark energy is a cosmological
constant, which is the primary goal of the study of dark energy, the constant EOS
 is the most direct form. Thus almost all cosmological observations data provide their
constraint on the constant $w$. By far its value is found to be very close to $w=-1$ and the
latest limit given by WMAP together with BAO and $H_0$ is $w=-1.10\pm0.14$ (Komatsu
et al. 2010). For the time-varying dark energy EOS, the simplest parameterizations
is the first-order Taylor expansion $w^{foT}(z) = w_0 + w' \, z$ (Huterer and Turner, 2001; Weller
and Albrechet, 2002), where $w_0$ and $w'$ are the value and the first-order
derivative of dark energy EOS at present, respectively. Though this model was used in
SNe Ia data analysis, its divergence behavior at the large redshift $z\rightarrow \infty $
is, in principle, disfavored  and at the same time, prevents it from CMB data fitting. The so-called
CPL parameterization $w^{CPL}(z) = w_0 + w_z \; z/ (1+z)$, proposed by Chevallier and Polarski (2001)
and Linder (2003), overcomes this defect and has capacities to describe a variety of scalar field theories.
For recent low redshift $z<<1$, $w^{CPL}$ is equivalent to the 1-order Taylor expansion
$w^{foT}$ mentioned above, and for high redshift, its value approaches $w_0 + w_z$.  By now,
the CPL parameterization has been adopted widely for all kinds of cosmological observations.

There are other remarkable works on the parameterization of dark energy EOS. The logarithmic
expansion in z was proposed to approximate scalar field EOS of the quintessence models with
smoothly varying potential (Efstathiou 2000; Gerke and Efstathiou 2002).  The CPL-extended parametrization
was proposed through changing the power exponent of the denominator in the CPL model from $1$ to $2$
(Jassal, Bagla and Padmanabhan 2005). Under this model, dark energy
evolves rapidly at high redshift and  has the same EOS at the present epoch and the low redshift.
Several attempts have also been made in the past to parameterize dark energy EOS $w(z)$ with more
than two parameters. The kink approach utilizes four observables to parameterize dark energy
EOS: the present value $w(z=0)$, the early value $w(z\rightarrow \infty$), the scale factor at the
transition point $a_t$ and the width of the transition $\Delta$ (Bassett et al. 2003, Corasaniti
and Copeland 2003, Corasaniti et al. 2004). This model is very useful when studying the
rapid transition from decelaerated to accelerated expansion. Using the gold sample of SNe
Ia (Riess et al. 2004) and a matter density prior, It was compared for the fitting results from
the standard Taylor expansions of $w(z)$ and the Kink parameterization (Bassett, Corasaniti and Kunz 2004).
For the form of Taylor expansion, it was considered with four different expansion functions and three different
expansion orders, highest to second order. With the marginalised on-dimensional likelihoods
for the parameters, it has been demonstrated that expansion order is more important than parametrisation.
After illustrating the `maximised' limits on the redshift of dark energy EOS $w(z)$,
it was claimed that the standard two-parameter expansions artificially rule out the models
associated with rapid evolution of dark energy. When focusing on the observational
information at high redshift, a variation of the Kink parameterization was adopted by
introducing the mean value $<w(z)>$ of dark energy EOS as an explicit parameter (Pogosian et al. 2005).
Different from the two-parameter models by fixing the value of the EOS today and
at high redshift with a definite evolution process, three-parameter or four-parameter
models are more flexible and allow for both rapid and slow variations of $w(z)$.

In our past work(Zhang et al. 2009; Zhang and Wu 2010), we have present the constraint on the
time-varying dark energy from the splitting angle statistic of strong gravitational lenses,
combined with SNe Ia and BAO observations. In this paper, by taking the SNe Ia data(Hicken et al. 2009),
the baryonic acoustic oscillations(Eisenstein et al. 2007), the 7-year WMAP observations(Komatsu et al. 2009),
together with the Hubble constant $H_0$ (Riess et al. 2009), we shall investigate a set of
two-parameter models of dark energy EOS $w(z)$. We will mainly highlight two issues which have not
previously been illuminated. First, we pay attention to the detailed study on some interesting two-parameter models with two
kinds of boundary behaviors. Second, we shall carefully demonstrate that there exist two kinds of strongly correlative dark energy
parametrizations so that the two parameters are reduced to one parameter. Our paper is organized
as follows: The cosmological observation data used in our analysis are presented in Sect. 2.
In Sect. 3, we make a joint analysis on a set of two-parameter models proposed in our present paper. The conclusions are
given in the last section.

\section{Relevant Cosmological Observation Data}
\label{sec:obsdata}

In this section we describe four sets of different cosmological observation data used
in our parameterization analysis, namely SNe Ia, BAO, CMB and Hubble constant $H_0$, which are
the currently preferred probes of the expansion history of the universe. Besides the
well-controlled system uncertainties and high measurement precisions, their analysis formulas
are quite simple and time-saving. This is important for our numerical calculations when
considering the relevant cosmological parameters simultaneously.
The  $H(z)$ observations from  Stern et al. (2010) have also been investigated and their
additional constraining power is not notable, and we shall not include the $H(z)$ sample here.
As the dark energy affects the space-time geometry and the distance determination, the SNe Ia sample
and BAO measurements provide us the distance ladder at low redshift $z<1.5$ and are expected
to distinguish the varieties of dark energy EOS $w(z)$ in the later epoch, while the much higher
characteristic redshift $z\sim1090$ of CMB  shall have somewhat "integral" effect and at the same
time, give the limit on $w(z)$ in the early universe. Combining above three observations makes it
reasonable to parameterize the dark energy EOS $w(z)$ by its two special  limiting values $w(z=0)$
and $w(z\rightarrow \infty)$. On the other hand, including the measurement of Hubble constant
$H_0$ can help to break the parameter degeneracies in  BAO and CMB observations.

\subsection{Type Ia supernovae}

It is widely believed that SNe Ia has homogeneous intrinsic luminosity of peak magnitude
and thus the SNe Ia data provide the precise distance measurements at the cosmic scale.
The analysis of the distance modulus versus redshift relation of SNe Ia  shows us the direct evidence of the
existence of dark energy. In the last decades, many SNe Ia observations have been
done and the total number of SNe Ia sample increases quickly(e.g. Riess et al. 2004;
Wood-Vasey et al. 2007; Kessler et al. 2009). The SN Ia compilations are
often consist of high-redshift $(z \approx 0.5)$ data set and low-redshift $(z \approx 0.05)$
sample at the same time. Remarkably, Kowalski et al. (2008) provided the Union data set with a compilation of
307 SNe Ia discovered in different surveys. The heterogeneous nature of the data set have been
reflected and all SNe Ia sample are analyzed with the same analysis procedure. Then by adding
the CAF3 sample, Hicken et al. (2009) shown the Constitution set with $397$ SNe Ia, of which
all SNe Ia light curves are fitted by using the spectral-template-based fit method of Guy
et al. (2005) (also known as SALT). In this paper, we use this constitution set to constrain
the dark energy EOS.

In the flat universe, the Friedmann equation are given by
 \beq
 H(z)/H_0 &=&  \sqrt{ \Omega_M (1+z)^3 + (1 - \Omega_M) f(z)}
 \nonumber \\
f(z) &=&  {\rm exp}\left(3 \int_0^z {1+w(z')\over1+z'} dz' \right)
 \eeq with Hubble constant $H_0 = 100 \; h \;{\rm km \;s^{-1}\;
 Mpc^{-1}}$.  The influence of dark energy EOS $w(z)$ is focused on the dark energy density
$\Omega_{DE}(z)= (1 - \Omega_M) f(z)$. For theoretical calculations, the luminosity distance $d_L$ of
SNe Ia is defined as
 \beq
 d_L(z) = (1+z) \int^z_0 {dz' \over H(z')}
 \eeq
In the SNe Ia observations, it is traditional to provide the distance moduli $\mu$,
which is related to the above luminosity distance $d_L$ by
 \beq
 \mu=5\log d_L/{\rm Mpc}+25.
 \eeq
Due to the entire degeneracy of the Hubble constant $H_0$ and the absolute magnitude of SNe Ia, the
parameter $H_0$ is irrelevant for the SNe Ia data analysis. Thus we use the following  $\chi^2$ statistic
(Perivolaropoulos (2005); Nesseris and  Perivolaropoulos (2005))
 \beq \chi^2_{SNe}(\Omega_M,w) =A(\Omega_M,w)-{B^2(\Omega_M,w)\over
C}. \label{eq:snechi2}
 \eeq where
 \beq A(\Omega_M,w)&=&\sum_i {(\mu^O_i-\mu^T_i(z_i;\Omega_M, w))^2\over
 \sigma_i^2}, \nonumber \\
B(\Omega_M,w)&=&\sum_i{\mu^O_i-\mu^T_i(z_i;\Omega_M, w)\over
\sigma_i^2}, \nonumber \\
 C&=& \sum_i{1\over \sigma_i^2}.
 \eeq
Here subscript $i$ denotes the $i$th SNe Ia data and $\sigma_i$ is
the observed uncertainty. The $\mu^O$ and $\mu^T$ are the observed and
theoretical distance moduli, respectively. Note that the $\mu^T_i(z_i)$ is independent on
$H_0$ and given by
 \beq
 \mu^T_i(z_i) = 5 \log (H_0d_L(z_i))
 \eeq

\subsection{Baryon Acoustic Oscillations}

The observations of baryon acoustic oscillations allow to constrain the
distance-redshift relation at different epochs, which is continuously improved
by the Sloan Digital Sky Survey (SDSS; York et al. 2000; Eisenstein et al. 2005;
Davis et al. 2007). From the final set of galaxies observed using the
original SDSS target selection criteria together with the 2-degree field Galaxy Redshift
Survey data (Colless et al. 2003; Cole et al. 2005), Percival et al. (2010) released
the new distance measurements at the redshift $z=0.2$ and $0.35$.
This combined sample is composed of $893,319$ galaxies over $9100$ square degrees
in the redshift range $0$ to $0.5$. The BAO relevant distance measure is modeled
by the so-called ¡°volume distance¡±, $D_V(z) = [D^2_A(z) z / H(z)]^{1/3}$, with comoving angular
diameter distance $D_A(z) = \int_0^z dz'/H(z')$. Define the distance ratio
$d_z\equiv r_s(z_d)/D_V(z)$, where $z_d$ is the redshift of the baryon drag epoch, which
can be calculated from the fitting formula proposed by Eisenstein and Hu (1998)
 \beq
 z_d = {1291 (\Omega_mh^2)^{0.251}\over 1+0.659(\Omega_mh^2)^{0.828}}[1+b_1(\Omega_bh^2)^{b_2}]
 \eeq where
 \beq
 b_1 = 0.313 (\Omega_mh^2)^{-0.419}[1+0.607(\Omega_mh^2)^{0.674}], \;\;\; b_2 = 0.238(\Omega_mh^2)^{0.223}
 \eeq
The observed distance ratios are given as:
 \beq
 d_{0.2} = 0.1905 \pm 0.0061  , \;\;\;\; d_{0.35} = 0.1097 \pm 0.0036.
 \eeq
and the inverse covariance matrix is given by
\begin{equation}
  C_{BAO}^{-1} = \left(\begin{array}{cc}
    30124 & -17227 \\
   -17227 &  86977 \\
   \end{array}\right).
\end{equation}
Then the $\chi^2$ statistic is
 \beq
 \chi^2_{BAO} = (d^{O}_{z_i} - d^{T}_{z_i})(C^{-1})_{i,j}(d^{O}_{z_j} - d^{T}_{z_j}) \label{eq:baochi2}
 \eeq

\subsection{Cosmic Microwave Background}
The CMB radiation is the cooled remnant of the hot photon-electron plasma in the
decoupling epoch of our universe and its precise measurements are critical to the cosmology.
The WMAP satellite has been measuring CMB temperature and polarization anisotropies over
the full sky since $2001$. After $7$ years of observations, Komatsu et al. (2009) provided
the latest fit results of WMAP Distance Priors, including the acoustic scale $l_A$, the shift
parameter $R$, and the redshift of the decoupling epoch of photons $z_{dc}$. The first two
parameters capture most of the constraining power of the WMAP data for dark energy properties,
which are given by
 \beq
 l_A &=& {\pi D_A(z_{dc})\over r_s(z_{dc})} \nonumber \\
 R(z_{dc})&=& \sqrt{\Omega_m H_0^2}D_A(z_{dc}).
 \eeq
Here the comoving sound horizon size at the decoupling epoch $r_s(z_{dc})$ is given by (Eisenstein and Hu 1998)
 \beq
 r_s(z_{dc}) = {1\over \sqrt{3}} \int_{z_{dc}}^{\infty} {dz \over H(z)\sqrt{1+(3\Omega_b/4\Omega_{\gamma})/(1+z)} }
 \eeq
with $\Omega_b$ and  $\Omega_{\gamma}$ the present-day baryon and photon density parameters, respectively.
In this paper, we fix $\Omega_{\gamma} = 2.469 \times 10^{-5} h^{-2}$ (for $T_{cmb}=2.725K$), which is the
best-fit values given by the 7-year WMAP observations (Komatsu et al. 2009).
 For the decoupling redshift $z_{dc}$, we use the fitting formula of Hu and Sugiyama (1996)
 \beq
 z_{dc}=1048[1+0.00124(\Omega_b h^2)^{-0.738}][1+g_1(\Omega_m h^2)^{g_2}],
 \eeq
where
 \beq
g_1=\frac{0.0783(\Omega_b h^2)^{-0.238}}{1+39.5(\Omega_b
h^2)^{0.763}},\quad g_2=\frac{0.560}{1+21.1(\Omega_b h^2)^{1.81}}.
 \eeq

Using above WMAP distance priors, the CMB $\chi^2$ statistic is given by
 \beq
 \chi^2_{CMB} = (x^{O}_i-x^{T}_i)(C^{-1})_{ij}(x^{O}_j-x^{T}_j) \label{eq:cmbchi2}
 \eeq where $x^T_i=(l_A, R, z_{dc})$ are the values predicted by a model and
$x^{O}_i$ are the corresponding maximum-likelihood
observe values. The three distance priors obtained from the WMAP 7-year are given as
 \beq
 l_A = 302.09 \pm 0.76, \;\; R = 1.725 \pm 0.018, \;\;  z_{dc}= 1091.3 \pm 0.91.
 \eeq
These results are correlated and the inverse covariance matrix is given by
 \beq
(C^{-1})=\left(
  \begin{array}{ccc}
    2.305 & 29.698 & -1.333 \\
    29.698& 6825.27 & -113.180 \\
    -1.333& -113.180 &  3.414 \\
  \end{array}
\right).
 \eeq

\subsection{Hubble Constant $H_0$}
The Hubble constant $H_0$ means the present expansion rate of our universe
and is very important for determining the cosmological distances. Using the
three observations in the local universe including the "maser galaxy" NGC 4258,
the Cepheid variables and the SNe Ia, Riess et al. (2009) provided a precise
measurement result $H_0 = 74.2\pm 3.6 \, { \rm km \, s^{-1}\, Mpc^{-1}}$.
In their analysis, the NGC 4258, located at about $7.2{\rm Mpc}$ away,  plays as the anchor galaxy,
for the overall uncertainty of its geometric distance is very small, only $3\%$
by far(Herrnstein et al. 1999; Humphreys et al. 2008; Greenhill et al. 2009).
On the other hand, the $240$ Cepheid variables obtained with the Hubble Space Telescope
(Macri et al. 2006) are distributed across six recent hosts of SNe Ia
and the NGC 4258, allowing to directly calibrate the peak luminosities  of the
SNe Ia, which is crucial to constrain the Hubble constant $H_0$ through the SNe Ia data.
The authors had rejected those objects with $0.75$ mag above the best fitted
period-luminosity relation, a error value beyond the "normal" observation uncertainty,
and the Cepheid sample is reduced to $209$. In the SNe Ia data analysis, only 140 nearby SNe Ia
with the redshift $0.023<z<0.1$ (Hicken et al. 2009) are used and
then, utilizing the derivations of the scale factor $a$, the related luminosity distance
is expanded to the polynomial function of the redshift $z$, keeping highest to $z^2$ order.
Finally with the well-controlled systematic errors, the SHOES team
provided the determination of the Hubble constant $H_0$ to $\sim \, 5\%$ precision.
Then the statistic is simply given by
 \beq
 \chi^2_{Hub}(H_0) = {[H_0 - 74.2]^2 \over 3.6^2 }
 \label{eq:H0chi2}
 \eeq

\subsection{Joint Analysis of SNe Ia, BAO, CMB and $H_0$ Data }
For the four independent observations, the likelihood function of a joint
analysis is just given by
 \beq
  L &=& L_{\rm SNe} \times L_{\rm BAO} \;\; \times L_{\rm CMB}  \;\; \times L_{\rm Hub}   \nonumber \\
    &=& \exp[-(\chi_{\rm SNe}^2+\chi_{\rm BAO}^2+\chi_{\rm CMB}^2+\chi_{\rm Hub}^2)/2].
 \eeq

\section{Constraints On Time-Varying Dark Energy Models }

In this section, we investigate our two-parameter models of dark energy EOS
from the joint analysis of  (SNe + BAO + BAO + $H_0$). The observational data
depend on $w(z)$ only through a complicated integral relation, which effectively
smooth out partial differences of different models. Consequently, to distinguish different
dark energy models from observational data is quite difficult.  In fact, though the use of more than
two parameters can offer flexible function form to test the dark energy EOS, the observational
constraints on those parameters are quite poor. Therefore, in this paper we focus our
efforts on  two-parameter models by carefully analysizing the low ($z\rightarrow0$) and high
redshift ($z\rightarrow \infty$) behaviors of dark energy EOS $w(z)$, while the certain function form
determines its evolution with time. Our two-parameter models are constructed with considering
three limitations: firstly, the dark energy EOS $w(z)$ has finite boundary values at $z=0$ and $z=\infty$; secondly, the $w(z)$
function has no more than one local extremum point in the range of redshift $z=(0, \infty)$;
thirdly, the basic bricks of the function are $z$, $1/(1+z)$(i.e. $a$)
and $z/(1+z)$(i.e. $(1-a)$). In  the data analysis, only two parameters are variable, and the possible relevant extra
parameter is fixed, which enlarges our model space. There is an alternative way to choose two parameters: assuming
that $w(z)$ can be approximated at low redshift $z<1$ and use the first two terms of its Taylor
series about $z=0$ (Copeland et al. 2006).  But when considering data from the CMB, the high redshift
limit becomes important and our choice is reasonable.

\subsection{Two-Parameter Dark Energy Models}

Though the certain knowledge of $w(z)$ is still lacked and the dark energy modeling space is
infinite dimensional, our aim is to find some interesting modeling forms and study the corresponding
behaviors of dark energy. We try different varieties of function forms with paying much attentions
to keep the models as simple as possible. According to the boundary behaviors of the EOS $w(z)$ at $z=0$ and $z=\infty$
and its local extremum property in the range of redshift $z = (0, \infty)$, we classify eight parameterations of dark
energy EOS $w(z)$ into two types of dark energy models, where three of the models have no extremum point and behave
analogously to the CPL model at two redshift boundaries $z=0$ and $z=\infty$. For other five models, they all have
one local extremum point with the same asymptotic values at the low and high redshift limitations.

Let us first consider three models which are analogous to the CPL model and classified as the type I models:
 \beq
{\rm Ia \;\;\;} :   w(z) &=& w_0 + w_1 z/(1+z)[1+1/(1+z)]  \label{eos:cpl3} \\
{\rm Ib \;\;\;} :   w(z) &=& w_0 + w_1 z {\rm ln}[(1+z)/z] \label{eos:cpl2} \\
{\rm Ic \;\;\;} :   w(z) &=& w_0 + w_1 (1 - {\rm ln}(1+z)/z)  \label{eos:cpl1} \\
{\rm CPL}:          w(z) &=& w_0 + w_1 z/(1+z)
 \eeq Just like CPL model, the three Type I models are all monotonic functions of redshift $z$ and have the
similar  boundary behaviors: $w(z=0) = w_0$ and $w(z \rightarrow \infty) = w_0 + w_1$.
Thus their variety in the whole domain is $|w_1|$, which implies a somewhat slow evolution of dark energy.
The dark energy EOS $w(z)$ of the type I models have no local extremum point and its evolution with time is simple.
In fact, the cosmological constant or the vacuum energy is the simplest model with the fixed EOS value $w\equiv-1$, and then
the constant model with a certain value unchanging with time.

As a simple extension to the type I and CPL models, we are now considering models in which there exist
one local extremum point for the dark energy EOS. In this case, the EOS $w(z)$ is no longer monotonic
and its derivative cross zero value in the past time. Such models are classified as type II models in our present considerations.
For the evolution pattern of the dark energy, we pay much attention to the logarithmic function forms.
As it is known that the dark energy EOS of some scalar field models can be approximated by the logarithmic function
of redshift $z$, and on the other hand, the quantum corrections of field theory are often given by the logarithmic
forms. Therefore, the  dark energy EOS could be related to the logarithmic function of the scale factor $a$.
By using the logarithmic form of $a=1/(1+z)$ and $(1-a)=z/(1+z)$, we can  construct the following four kinds of models
 \beq
{\rm IIa}:   w(z) &=& w_0  + w_1 [1/(1+z)]^{\alpha}  {\rm ln}[1/(1+z)]^{\alpha} \label{eos:log0} \\
{\rm IIb}:   w(z) &=& w_0  + w_1 [z/(1+z)]^{\alpha}  {\rm ln}[z/(1+z)]^{\alpha} \label{eos:log1} \\
{\rm IIc}:   w(z) &=& w_0  + w_1 {\rm ln}(1+z)/z^{\alpha} \label{eos:log2}\\
{\rm IId}:   w(z) &=& w_0  + w_1 z^{\alpha} {\rm ln}[(1+z)/z] \label{eos:log3}
  \eeq where ${\alpha}$ is an extra parameter and fixed in the data analysis. Note that Model Ic can be obtained
  through the $(w_0, w_1)$ redefinition of the model IIc, and the model Ib is the special case
of model IId with ${\alpha} = 1$. For ${\alpha} = 0$, the model IIa and model IIb are reduced to the constant model,
and the model IIc and model IId are reduced to the simple models with ${\rm ln(a)}$ and ${\rm ln(1-a)}$,
 which are divergent at boundaries $a=0$ and $a=1$ or $z=\infty$ and $z=0$. Therefore,  the case of ${\alpha}=0$
  will not be discussed here.  For certain values of ${\alpha}$, all models have one local extremum
in the range $z=0$ to $z\rightarrow \infty$, and its position is determined by the parameter ${\alpha}$.
The boundary behaviors of the models at high and low redshift depend on the parameter ${\alpha}$.
According to different choices of  ${\alpha}$, there are three boundary values: $w_0$, $w_0+w_1$ and $\infty$.
In this work, we are only interested in the former two cases. For comparison, we also present one  model without
logarithmic function form
 \beq
{\rm IIe}:  w(z) &=& w_0  + w_1 z/(1+z)^{\alpha} \label{eos:nl1}
  \eeq where ${\alpha}$ is an extra parameter and will be fixed in the data analysis. For ${\alpha}=0$,
the model IIe is reduced to $w(z) = w_0 + w_1 z$, which is divergent when $z \rightarrow \infty$ and
will not be considered in this work. The corresponding local extremum property is clear that
the model IIe  has one local extremum point for $\alpha > 1$. With different choices of ${\alpha}$,
there are three boundary values: $w_0$, $w_0+w_1$ and $\infty$. For the cases  ${\alpha}=1$ and ${\alpha}=2$, it was studied
in literature (Jassal, Bagla and Padmanabhan 2005). Here we shall consider $\alpha$ as an arbitrary parameter to be fixed.
In the data analysis late on, we will show that the local extremum property and the boundary
behavior together determine the pattern of $w(z)$ for its allowed region.

It is interesting to find in the joint analysis that there exist two strongly correlative parameterations
 \beq
{\rm A}:  w(z) &=& w_0  + w_1  z^{4/9} {\rm ln}[(1+z)/z] \label{eos:cor1}  \\
{\rm B}:  w(z) &=& w_0  + w_1  z^{-5/6} {\rm ln}(1+z), \label{eos:cor2}
  \eeq  which are the special cases of model IId and model IIc with ${\alpha}=4/9$ and $5/6$, respectively.
In these two cases, the resulting $w_0$ and $w_1$ have almost linear negative correlations with $w_1 = 0$ at $w_0=-1$.

The boundary behavior at low redshift $z \rightarrow 0$ and high redshift $z \rightarrow \infty$
for all models are summarized in the table \ref{tab:bound}. For the three type I models,
the  boundary values of dark energy EOS $w(z)$ are similar with the CPL model: $w(z \rightarrow 0) \rightarrow w_0$ and
$w(z \rightarrow  \infty) \rightarrow (w_0+w_1)$. As type I models are  monotonic function, their
range is just the inteval $w_0$ and $(w_0+w_1)$. For the four models with the logarithmic form,
the boundary behavior relies on the value of $\alpha$. For models IIa and IIb,
their boundary values at the low and high redshift are both $w_0$ when ${\alpha} \geq 0$. For models IIc an IId, their
boundary values are both $w_0$ for $0<\alpha<1$. For the case of ${\alpha}=1$, the model IIc has the low redshift
value $(w_0 + w_1)$ and high redshift value $w_0$. For model IIe, the two boundary values are both $w_0$ when $\alpha > 1$.
For the two strongly correlative models, their boundary values are also both $w_0$.

The local extremum properties  of all models have been summarized in Table \ref{tab:extremum}.
Though the interested model domain in this work is just for the range with $z \geq 0$, here the redshift
relating to the future with $-1<z<0$ is also included and the local extremum point is given. For model IIa or model IIb, there is one local extremum point located at $z^{*}= e^{1/{\alpha}}-1$ or $z^{*}= 1/(e^{1/{\alpha}}-1)$ for $\alpha>0$. Model IIa has one local extremum value in future for $\alpha<0$. The extremum properties of model IIc is somewhat complex. One local extremum point is located at $z^{*}/(1+z^{*})={\alpha}{\rm ln}(1+z^{*})$ for $0<\alpha<1$, and for the case $\alpha>1$ one local extremum point can be
found in the redshift range $-1<z<0$. Model IId has one local extremum value at $1/(1+z^{*})={\alpha}{\rm ln}[(1+z^{*})/z^{*}]$
for $0<{\alpha}<1$. For the model IIe without logarithmic function, the local extremum point is found to be at
$z^{*}= 1/({\alpha}-1)$ for ${\alpha}>1$. For the two strongly correlative models,
their local extremum positions can be obtained from above general cases with specific values of $\alpha$, the corresponding
values of the redshift are found to be  $z^* =0.17$ for model A and $z^* =0.46$ for model B.

The different choice of the extra parameter ${\alpha}$  will change the boundary values and local extremum position
of dark energy EOS. For the type II dark energy models, we are interested in the cases
with finite boundary values and one local extremum point at $z>0$. From the summaries given in
two tables, the interesting ranges for the extra parameter $\alpha$ are found to be: ${\alpha}>0$ for model IIa and mode
IIb, $1 < {\alpha} \leq1$ for model IIc and model IId, and ${\alpha} \geq 1$ for model IIe. In these cases, the low redshift
and high redshift values for all the type II five models are equal: $w(z\rightarrow 0) = w(z\rightarrow \infty) = w_0$.

\begin{table}
\begin{tabular}{|c|l|c|c|c|}
 \hline  {\rule[4pt]{0mm}{10pt}  model}& \hspace{0.5cm} equation  &  {$\alpha$}  & $z \rightarrow 0$ &   $z \rightarrow \infty$ \\ \hline \hline\rule[4pt]{0mm}{10pt}

CPL    & $w(z) = w_0 + w_1 z/(1+z)$ &  &  $w_0$ &  $w_0 + w_1$      \\ \cline{1-5}\rule[4pt]{0mm}{10pt}

Ia     & $w(z) = w_0 + w_1 z/(1+z)[1+1/(1+z)]$ &  &  $w_0$ &  $w_0 + w_1$      \\ \cline{1-5}\rule[4pt]{0mm}{10pt}

Ib     & $w(z) = w_0 + w_1 z {\rm ln}[(1+z)/z]$ &  &  $w_0$ &  $w_0 + w_1$      \\ \cline{1-5}\rule[4pt]{0mm}{10pt}

Ic     & $w(z) = w_0 + w_1 (1 - {\rm ln}(1+z)/z)$ &  &  $w_0$ &  $w_0 + w_1$     \\ \hline \hline\rule[4pt]{0mm}{10pt}

\multirow{2}{*}{IIa}   & \multirow{2}{*}{$w(z) = w_0  + w_1 [1/(1+z)]^{\alpha}  {\rm ln}(1/(1+z))^{\alpha}$} & ${\alpha}\geq 0$ & $w_0$ &  $w_0$     \\\cline{3-5}\rule[4pt]{0mm}{10pt}

      &   & ${\alpha}<0$ & $w_0$ &  $\infty$     \\ \cline{1-5}\rule[4pt]{0mm}{10pt}

\multirow{2}{*}{IIb}   & \multirow{2}{*}{$w(z) = w_0  + w_1 [z/(1+z)]^{\alpha}  {\rm ln}(z/(1+z))^{\alpha}$}  & ${\alpha}\geq 0$ & $w_0$ &  $w_0$     \\ \cline{3-5}\rule[4pt]{0mm}{10pt}

      &   & ${\alpha}<0$ & $\infty$ &  $w_0$    \\ \cline{1-5}\rule[4pt]{0mm}{10pt}

\multirow{4}{*}{IIc}      & \multirow{4}{*}{$w(z) = w_0  + w_1 {\rm ln}(1+z)/z^{\alpha}$} & ${\alpha}>1$ & $\infty$ &  $w_0$     \\ \cline{3-5}\rule[4pt]{0mm}{10pt}

      &   & ${\alpha}=1$ & $w_0 + w_1$ &  $w_0$     \\ \cline{3-5}\rule[4pt]{0mm}{10pt}

      &   & $0< {\alpha}<1$ & $w_0$  &  $w_0$     \\ \cline{3-5}\rule[4pt]{0mm}{10pt}

      &   & ${\alpha}\leq0$ & $w_0$ &  $\infty$     \\ \cline{1-5}\rule[4pt]{0mm}{10pt}

\multirow{4}{*}{ IId }    & \multirow{4}{*}{$w(z) = w_0  + w_1 z^{\alpha} {\rm ln}[(1+z)/z]$} & ${\alpha}>1$ &  $w_0$ &  $\infty$   \\ \cline{3-5}\rule[4pt]{0mm}{10pt}

      &   & ${\alpha}=1$ & $w_0$ &  $w_0 + w_1$   \\ \cline{3-5}\rule[4pt]{0mm}{10pt}

      &   & $0< {\alpha}<1$ & $w_0$  &  $w_0$     \\ \cline{3-5}\rule[4pt]{0mm}{10pt}

      &   & ${\alpha}\leq0$ & $\infty$ &  $w_0$   \\ \cline{1-5}\rule[4pt]{0mm}{10pt}

\multirow{3}{*}{IIe }& \multirow{3}{*}{$w(z) = w_0  + w_1 z/(1+z)^{\alpha}$} & ${\alpha}>1$ &  $w_0$  &  $w_0$  \\  \cline{3-5}\rule[4pt]{0mm}{10pt}

      &   & ${\alpha}=1$ & $w_0$ &  $w_0 + w_1$   \\ \cline{3-5}\rule[4pt]{0mm}{10pt}

      &   & ${\alpha}<1$ & $w_0$ &  $\infty$   \\ \hline \hline\rule[4pt]{0mm}{10pt}

A    &  $w(z) = w_0  + w_1  z^{4/9} {\rm ln}[(1+z)/z]$ & {$\alpha$}=4/9 & $w_0$  &  $w_0$ \\ \cline{1-5}\rule[4pt]{0mm}{10pt}

B    &  $w(z) = w_0  + w_1  {\rm ln}(1+z)/z^{5/6}$ & {$\alpha$}=5/6 & $w_0$  &  $w_0$
\\ \hline
\end{tabular}
\caption{The boundary behaviors of all models}\label{tab:bound}
\end{table}

\begin{table}
\begin{tabular}{|c|c|c|}
 \hline  {\rule[4pt]{0mm}{10pt}  model}   &  {$\alpha$}   & extremum point: $z^{*}$ \\ \hline \hline\rule[4pt]{0mm}{10pt}

CPL     &  &   no       \\ \cline{1-3}\rule[4pt]{0mm}{10pt}

Ia      &  &  no     \\ \cline{1-3}\rule[4pt]{0mm}{10pt}

Ib      &  &  no      \\ \cline{1-3}\rule[4pt]{0mm}{10pt}

Ic      &  &  no       \\ \hline \hline\rule[4pt]{0mm}{10pt}

\multirow{2}{*}{IIa}     & ${\alpha} > 0$ & $z^{*}= e^{1/{\alpha}}-1$    \\\cline{2-3}\rule[4pt]{0mm}{10pt}

        & ${\alpha} < 0$ &   in future   \\ \cline{1-3}\rule[4pt]{0mm}{10pt}

\multirow{2}{*}{IIb}     & ${\alpha} > 0$ & $z^{*}= 1/(e^{1/{\alpha}}-1)$    \\ \cline{2-3}\rule[4pt]{0mm}{10pt}

        & ${\alpha} < 0$ & no    \\ \cline{1-3}\rule[4pt]{0mm}{10pt}

\multirow{5}{*}{IIc}        & $0<{\alpha}<1$ & $z^{*}/(1+z^{*})={\alpha}{\rm ln}(1+z^{*}) $     \\ \cline{2-3}\rule[4pt]{0mm}{10pt}

         & ${\alpha} = 1$ &  no      \\ \cline{2-3}\rule[4pt]{0mm}{10pt}

         & ${\alpha} > 1$ &  in future      \\ \cline{2-3}\rule[4pt]{0mm}{10pt}

         & $0<{\alpha}<1$ &  \multirow{2}{*}{ $z^{*} = 0$ \; or \; no } \\ \cline{2-2}\rule[4pt]{0mm}{10pt}

         & ${\alpha}<0$ &    \\ \cline{1-3}\rule[4pt]{0mm}{10pt}

\multirow{2}{*}{IId}        & $0<{\alpha}<1$ & $1/(1+z^{*})={\alpha}{\rm ln}[(1+z^{*})/z^{*}] $     \\ \cline{2-3}\rule[4pt]{0mm}{10pt}

          & ${\alpha}\geq1 \; {\rm or}  \; {\alpha}<0$ & no   \\   \cline{1-3}\rule[4pt]{0mm}{10pt}

\multirow{3}{*}{IIe}    & ${\alpha}>1$ &  $z^{*}= 1/({\alpha}-1)$  \\ \cline{2-3}\rule[4pt]{0mm}{10pt}

        & $0<{\alpha}\leq1$ & no   \\ \cline{2-3}\rule[4pt]{0mm}{10pt}

        & ${\alpha}<0$ & in future   \\  \hline \hline \rule[4pt]{0mm}{10pt}

A       & ${\alpha}=4/9$ & $z^{*}= 0.17$    \\ \cline{1-3}\rule[4pt]{0mm}{10pt}

B       & ${\alpha}=5/6$ & $z^{*}= 0.46$
\\ \hline
\end{tabular}
\caption{The local extremum point properties of all models} \label{tab:extremum}
\end{table}

\subsection{Results From Latest Observations}

\begin{table}
\begin{tabular}{|c|c|c|c|c|}
 \hline \multirow{2}{*}{\rule[4pt]{0mm}{10pt}  model}&   \multirow{2}{*}{ {$\alpha$} } &
 \multicolumn{2}{|c|}{\;\;\;\;($w_0,w_1, \Omega_M,h,100\Omega_bh^2$)} & \multirow{2}{*}{$(w_0, w_1)$}  \\\cline{3-4}\rule[4pt]{0mm}{10pt}

 & &  best-fit 5 parameters & $\chi^2_{min}$ & \\ \hline \hline\rule[4pt]{0mm}{10pt}

CPL      &  & (-0.97, -0.19),\;( 0.28, 0.70, 2.24) &  469.045   &  $(-0.97^{+0.22}_{-0.18}, -0.15^{+0.85}_{-1.33})$  \\ \cline{1-5}\rule[4pt]{0mm}{10pt}

Ia       &  & (-0.97, -0.15),\;( 0.28, 0.70, 2.24)  &  469.027  &  $(-0.96^{+0.26}_{-0.21}, -0.12^{+0.61}_{-0.89})$  \\ \cline{1-5}\rule[4pt]{0mm}{10pt}

Ib       &  & (-0.95, -0.14),\;( 0.28, 0.70, 2.24) &  468.992   &  $(-0.93^{+0.34}_{-0.28}, -0.19^{+0.76}_{-1.05})$  \\ \cline{1-5}\rule[4pt]{0mm}{10pt}

Ic       &  & (-0.93, -0.19),\;( 0.28, 0.70, 2.24) &  468.959   &  $(-0.97^{+0.19}_{-0.17}, +0.22^{+0.81}_{-2.59})$  \\ \hline  \hline\rule[4pt]{0mm}{10pt}

\multirow{4}{*}{IIa}      & 1 & (-0.94, 0.33),\;( 0.28, 0.70, 2.23) &  468.971   &  $(-0.95^{+0.27}_{-0.23}, +0.32^{+1.88}_{-1.41})$  \\\cline{2-5}\rule[4pt]{0mm}{10pt}

         & 2 & (-0.86, 0.63),\;( 0.28, 0.70, 2.23) &  468.732   &  $(-0.86^{+0.38}_{-0.36}, +0.64^{+1.74}_{-1.61})$  \\\cline{2-5}\rule[4pt]{0mm}{10pt}

         & 3 & (-0.66, 1.33),\;( 0.29, 0.69, 2.24) &  468.199   &  $(-0.66^{+0.55}_{-0.51}, +1.32^{+2.29}_{-2.05})$  \\ \cline{2-5}\rule[4pt]{0mm}{10pt}

         & 4 & (-0.57, 1.61),\;( 0.29, 0.69, 2.24) &  467.908   &  $(-0.57^{+0.40}_{-0.63}, +1.58^{+1.66}_{-2.31})$ \\ \cline{1-5}\rule[4pt]{0mm}{10pt}

\multirow{4}{*}{IIb}      & 1 & (-0.77, 0.87),\;( 0.28, 0.70, 2.23) &  468.644   &  $(-0.77^{+0.59}_{-0.56}, +0.87^{+2.28}_{-1.10})$ \\ \cline{2-5}\rule[4pt]{0mm}{10pt}

         & 2 & (-0.97, 0.22),\;( 0.28, 0.70, 2.24) &  469.023   &  $(-0.96^{+0.17}_{-0.18}, +0.28^{+0.66}_{-1.33})$  \\ \cline{2-5}\rule[4pt]{0mm}{10pt}

         & 3 & (-0.99, 0.03),\;( 0.28, 0.70, 2.24) &  469.077    &  $(-0.99^{+0.12}_{-0.12}, +0.02^{+2.42}_{-1.43})$ \\ \cline{2-5}\rule[4pt]{0mm}{10pt}

         & 4 & (-1.00, -0.37),\;( 0.28, 0.70, 2.24) &  469.074   &  $(-0.99^{+0.09}_{-0.11}, -0.04^{+3.73}_{-1.79})$  \\ \cline{1-5}\rule[4pt]{0mm}{10pt}

\multirow{4}{*}{IIc}      & 1/5 & (-0.96, -0.11),\;( 0.28, 0.70, 2.24) &  469.026  &  $(-0.97^{+0.25}_{-0.19}, -0.10^{+0.59}_{-1.02})$   \\ \cline{2-5}\rule[4pt]{0mm}{10pt}

          & 1/3 & (-0.95, -0.15),\;( 0.28, 0.70, 2.24) &  469.000  &  $(-0.94^{+0.30}_{-0.26}, -0.16^{+0.72}_{-1.05})$  \\ \cline{2-5}\rule[4pt]{0mm}{10pt}

         & 1/2 & (-0.90, -0.23),\;( 0.28, 0.70, 2.24) &  468.949  &  $(-0.91^{+0.45}_{-0.37}, -0.22^{+0.89}_{-1.21})$ \\ \cline{2-5}\rule[4pt]{0mm}{10pt}

         & 2/3 & (-0.79, -0.41),\;( 0.28, 0.70, 2.23) &  468.862   &  $(-0.80^{+0.75}_{-0.64}, -0.38^{+1.24}_{-1.53})$  \\ \cline{1-5}\rule[4pt]{0mm}{10pt}

\multirow{4}{*}{IId}      & 1/5 & (-1.14,  0.12),\;( 0.28, 0.70, 2.24) &  469.956  &  $(-1.13^{+0.55}_{-0.82}, +0.10^{+0.59}_{-0.42})$  \\ \cline{2-5}\rule[4pt]{0mm}{10pt}

        & 1/3 & (-1.10,  0.11),\;( 0.28, 0.70, 2.24) &  469.053  &  $(-0.97^{+0.70}_{-1.65}, -0.02^{+1.58}_{-0.71})$  \\ \cline{2-5}\rule[4pt]{0mm}{10pt}

         & 2/3 & (-0.67, -0.58),\;( 0.28, 0.70, 2.23) &  468.719   &  $(-0.68^{+0.88}_{-0.80}, -0.56^{+1.41}_{-1.59})$  \\ \cline{2-5}\rule[4pt]{0mm}{10pt}

         & 4/5 & (-0.84, -0.33),\;( 0.28, 0.70, 2.23) &  468.857   &  $(-0.87^{+0.57}_{-0.45}, -0.28^{+0.96}_{-1.27})$\\ \cline{1-5}\rule[4pt]{0mm}{10pt}

\multirow{4}{*}{IIe}    & 5/4 & (-0.96, -0.22),\;( 0.28, 0.70, 2.24) &  469.006  &  $(-0.96^{+0.24}_{-0.21}, -0.21^{+1.08}_{-1.49})$   \\ \cline{2-5}\rule[4pt]{0mm}{10pt}

        & 4/3   & (-0.95, -0.25),\;( 0.28, 0.70, 2.24) &  468.997  &  $(-0.95^{+0.25}_{-0.21}, -0.24^{+1.16}_{-1.66})$  \\ \cline{2-5}\rule[4pt]{0mm}{10pt}

        & 3/2 & (-0.95, -0.32),\;( 0.28, 0.70, 2.24) &  468.977  &  $(-0.94^{+0.26}_{-0.23}, -0.31^{+1.38}_{-1.84})$   \\ \cline{2-5}\rule[4pt]{0mm}{10pt}

        & 7/4   & (-0.93, -0.45),\;( 0.28, 0.70, 2.24) &  468.939  &  $(-0.93^{+0.28}_{-0.26}, -0.44^{+1.66}_{-2.06})$   \\  \hline \hline\rule[4pt]{0mm}{10pt}

A      & 4/9 & (-0.57, -0.51),\;( 0.28, 0.69, 2.24) &  468.947  &  $(-0.57^{+0.85}_{-2.38}, -0.52^{+2.88}_{-1.09})$ \\ \cline{1-5}\rule[4pt]{0mm}{10pt}

B      & 5/6 & (0.16, -1.74),\;( 0.29, 0.69, 2.24) &   468.525  &  $(+0.15^{+0.78}_{-2.28}, -1.72^{+3.43}_{-1.30})$
\\ \hline
\end{tabular}
\caption{Based on the joint analysis of (SNe + BAO + CMB + $H_0$), the best-fit results of five parameters
($w_0,w_1, \Omega_M,h,100\Omega_bh^2$) and the corresponding minimum $\chi^2_{min}$, as well as the
constraint results of $(w_0, w_1)$ with other three cosmological parameters marginalized.}\label{tab:bestf}
\end{table}

Based on the above analysis, we present in table \ref{tab:bestf} the best-fit results of five parameters
($w_0,w_1, \Omega_M,h,100\Omega_bh^2$) and the corresponding minimum $\chi^2_{min}$,
as well as the constraint results of $(w_0, w_1)$ with other three cosmological parameters marginalized
from the joint analysis of (SNe + BAO + CMB + $H_0$).  It can been seen that all
$\chi^2_{min}$ are close to $\chi^2_{min} \approx 469$, which indicates that the current observational data
can not explicitly distinguish different two-parameters models from the $\chi^2_{min}$ analysis. For all models, the
constraint results for three of the cosmological parameters ($\Omega_M,h,100\Omega_bh^2$)
are close to (0.28, 0.70, 2.24), though the best-fit results for
the two parameters $w_0$ and $w_1$ are quite different. For models IIa, IIb, and IIe, the best-fit
values of $(w_0, w_1)$ vary regularly according to the change of the extra parameter ${\alpha}$.
For models IIc and IId, there is a turnaround point in the range $2/3<z<1$ and $1/3<z<1/2$, respectively.

\begin{figure}
\epsscale{0.9} \plotone{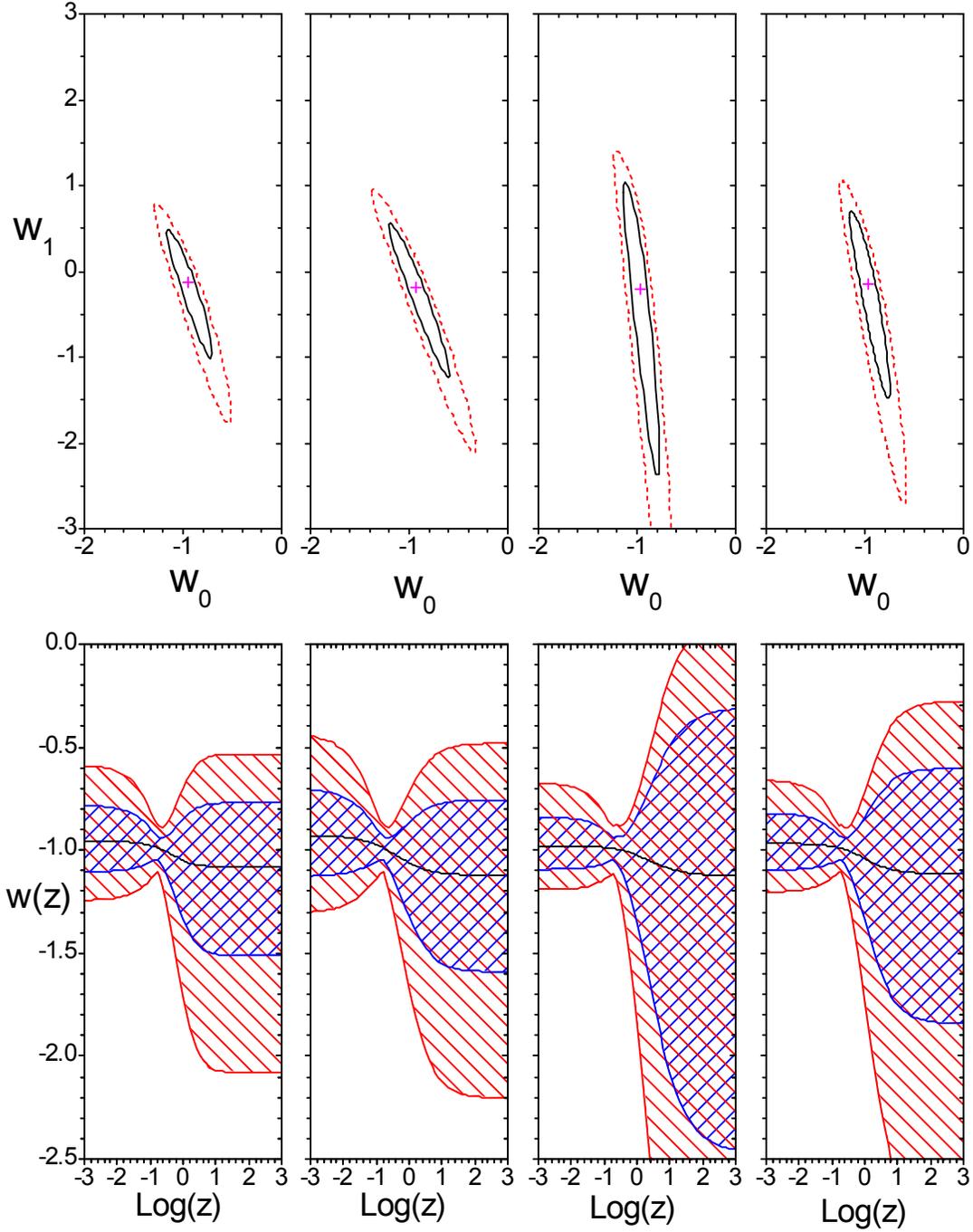}
\figcaption{ The $68\%$ and $95\%$ confidence contours of four models: $w(z) = w_0 + w_1 z/(1+z)[2-z/(1+z)]$,
$w(z) = w_0 + w_1 z {\rm ln}[(1+z)/z]$, $w(z) = w_0 + w_1 (1 - {\rm ln}(1+z)/z)$,
$w(z) = w_0 + w_1 z/(1+z)$, from left to right. The solid lines in bottom panels show the best fit results.
The crosshairs in top panels mark the best-fit points ($w_0, w_1)=(-0.96, -0.12)$, $(-0.93,-0.19)$, $(-0.97, -0.22)$
and $(-0.97, -0.15)$, from left to right.
 \label{fig:cpl}}
 \end{figure}

Let us first discuss the constraints for the three type I models, and compare them with the CPL model. In
figure \ref{fig:cpl}, we show the constraint results on $(w_0, w_1)$ and $w(z)$ from the joint
analysis of (SNe + BAO + CMB + $H_0$). The crosshairs in top panels mark the best-fit points
 ($w_0, w_1)=(-0.96, -0.12)$, $(-0.93,-0.19)$, $(-0.97, -0.22)$ and $(-0.97, -0.15)$,
from left to right. In bottom panels, the best fit results are shown as
solid lines. Clearly, these panels display the following common features. Firstly,
it shows negative dependence between the parameters $w_0$ and $w_1$, which indicates why
the area of allowed region for $w(z)$ below the best-fit line is larger than the one above the best-fit line.
Secondly, around $z\sim 0.2$, the  allowed region for $w(z)$  has a narrow node,
at which the uncertainty of $w(z)$ becomes minimum due to correlations of the parameters
$w_0$ and $w_1$ (see, e.g., Linder and Huterer 2005; Albrecht et al. 2006). Thirdly,
the constraining power at high redshift $z>>1$ is weaker than that at low redshift $z\sim1$.
The reason is that there are more observational data at low redshift
than at high redshift and the relative low density of dark energy at high redshift makes the data insensitive
to the model variety. Finally, the resulting best-fit $w(z)$ for all the modles is similar with a value around $w\simeq -1$.
It can be seen that the constraining power on $w(z)$ is different for various models and the model Ia
considered in our present paper leads to the best constraint results.

\begin{figure}
\epsscale{0.9} \plotone{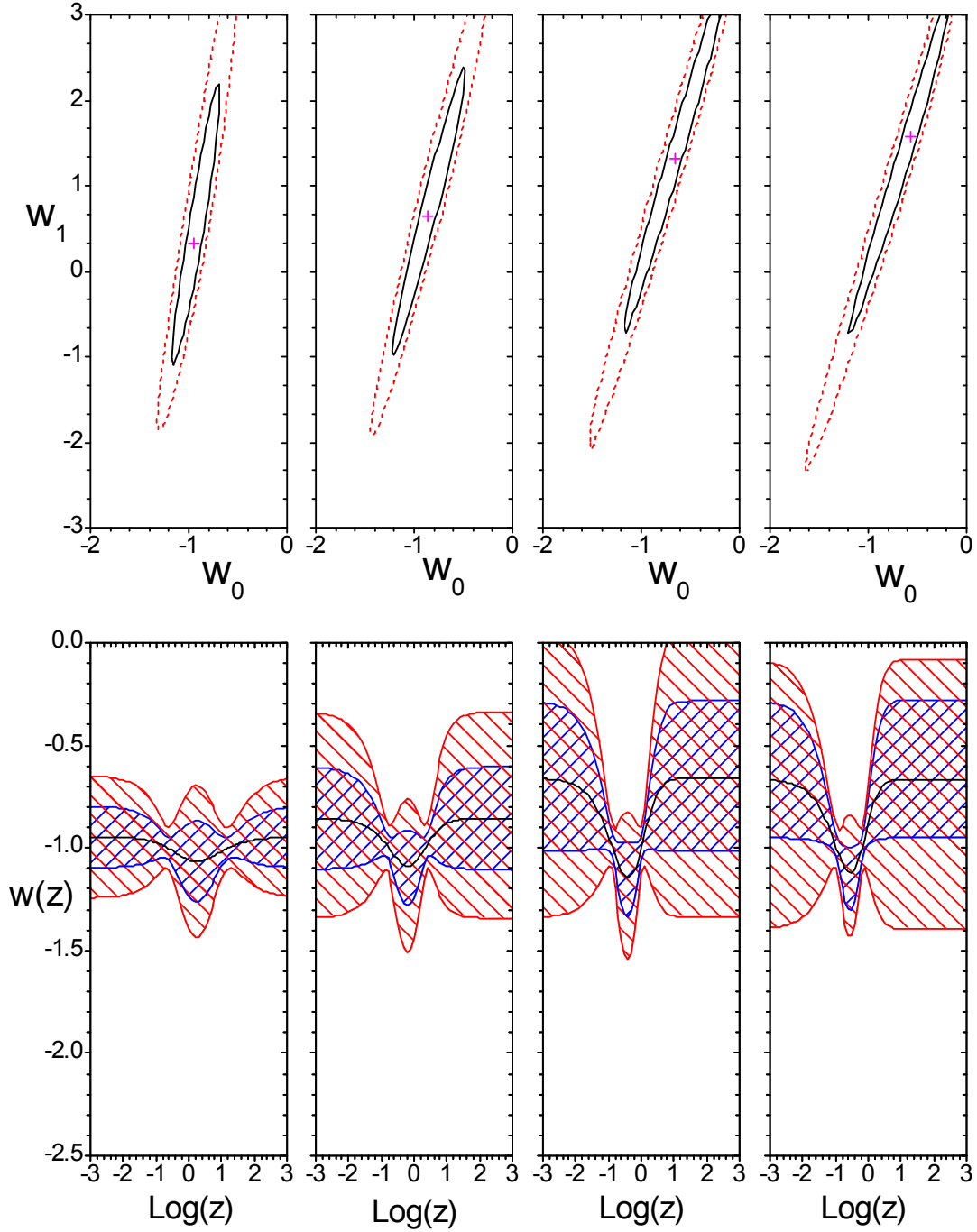}
\figcaption{  The $68\%$ and $95\%$ confidence contours of the logarithmic form model
$w(z) = w_0  + w_1 [1/(1+z)]^{\alpha}  {\rm ln}(1/(1+z))^{\alpha}$ with ${\alpha}=1$,$2$,$3$ and $4$, from left to right.
The solid lines in bottoms panels show the best fit results.
The crosshairs in top panels mark the best-fit points ($w_0, w_1)=(-0.95, 0.32)$, $(-0.86, 0.64)$, $(-0.66,1.32)$
and $(-0.57, 1.58)$, from left to right. \label{fig:log0}}
 \end{figure}

We now consider the type II models based on the joint analysis of four observations (SNe + BAO + CMB + $H_0$). Figure \ref{fig:log0} plots the $68\%$ and $95\%$ confidence contours of $(w_0, w_1)$ and $w(z)$
for the model IIa  with the given extra parameter ${\alpha} = 1, 2, 3$ and $4$, respectively.
The crosshairs in top panels mark the best-fit points ($w_0, w_1)=(-0.95, 0.32)$, $(-0.86, 0.64)$,
$(-0.66,1.32)$ and $(-0.57, 1.58)$, from left to right. The solid lines in bottom panels show
the best-fit results. It shows the following common features£º Firstly,
the two parameters $w_0$ and $w_1$ have positive correlations. Secondly, the confidence intervals of
$w_1$ is much bigger than that of $w_0$, which means that the observational data are less sensitive to $w_1$.
Thirdly, there exist two narrow nodes in the allowed regions and the node corresponding to the small redshift is near $z \sim 0.1$.
Such a feature is caused  by the local extremum point in the model IIa around $z = 1$.
Fourthly, the constraining power at low redshift $z<<1$ is similar to that at high redshift $z>>1$,
which is partially due to the same boundary behaviors of low redshift and high redshift:
$w(z \rightarrow 0) = w(z \rightarrow  \infty) \rightarrow w_0$.
When the parameter ${\alpha}$ is taken to be a larger value, the allowed region between two narrow nodes
goes smaller, the best-fit line curves down,  and the relative constraining power becomes weaker at both low redshift and high redshift.

\begin{figure}
\epsscale{0.9} \plotone{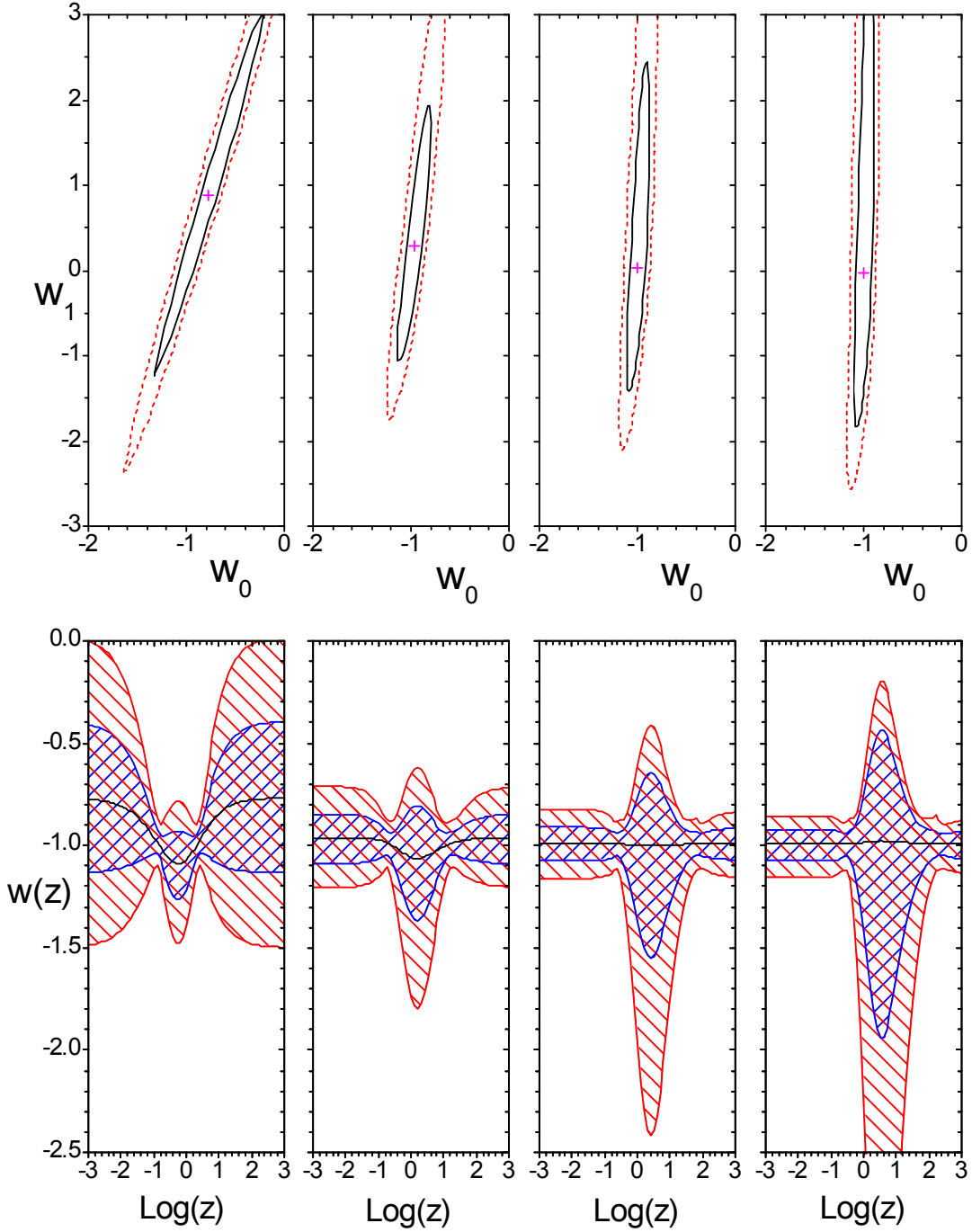}
\figcaption{  The $68\%$ and $95\%$ confidence contours of the logarithmic form model
$w(z) = w_0  + w_1 [z/(1+z)]^{\alpha}  {\rm ln}(z/(1+z))^{\alpha}$ with ${\alpha}=1$,$2$,$3$ and $4$, from left to right.
The solid lines in bottoms panels show the best fit results.
The crosshairs in top panels mark the best-fit points ($w_0, w_1)=(-0.77, 0.87)$, $(-0.96, 0.28)$, $(-0.99, 0.02)$
and $(-0.99, -0.04)$, from left to right. \label{fig:log1}}
 \end{figure}

We plot in figure \ref{fig:log1} the $68\%$ and $95\%$ confidence contours for the model IIb
with various values of parameter ${\alpha}=1$,$2$,$3$ and $4$. The crosshairs in top
panels mark the best-fit points ($w_0, w_1)=(-0.77, 0.87)$, $(-0.96, 0.28)$, $(-0.99, 0.02)$
and $(-0.99, -0.04)$, from left to right. The solid lines in bottom panels show the best-fit results.
It is not difficult to observe from the figure 3 the following  common characters:
Firstly, the allowed range of $w_1$ is much bigger than $w_0$ and
they have strong positive correlations. Secondly, there are two narrow nodes in the  allowed regions.
Thirdly, the constraining power at low redshift $z<<1$ is analogous to that at high redshift $z>>1$,
which can be explained by the same boundary behaviors of low redshift and high redshift:
$w(z \rightarrow 0) = w(z \rightarrow  \infty) \rightarrow w_0$.  For a larger value of parameter ${\alpha}$,
the allowed interval of $w(z)$ between two narrow nodes increases rapidly, and the constraint power gets stronger
for low and high redsfhits, which just opposites to the case in the model IIa. Especially,
the behavior of phantom-like dark energy model indicated by the case with ${\alpha} = 4$
is consistent with the current observations: the EOS $w(z)$ is close to $-1$ in both
low redshift $z<0.5$ and high redshift $z>100$ and decreases to  much negative values when $z \rightarrow 4$.

\begin{figure}
\epsscale{0.9} \plotone{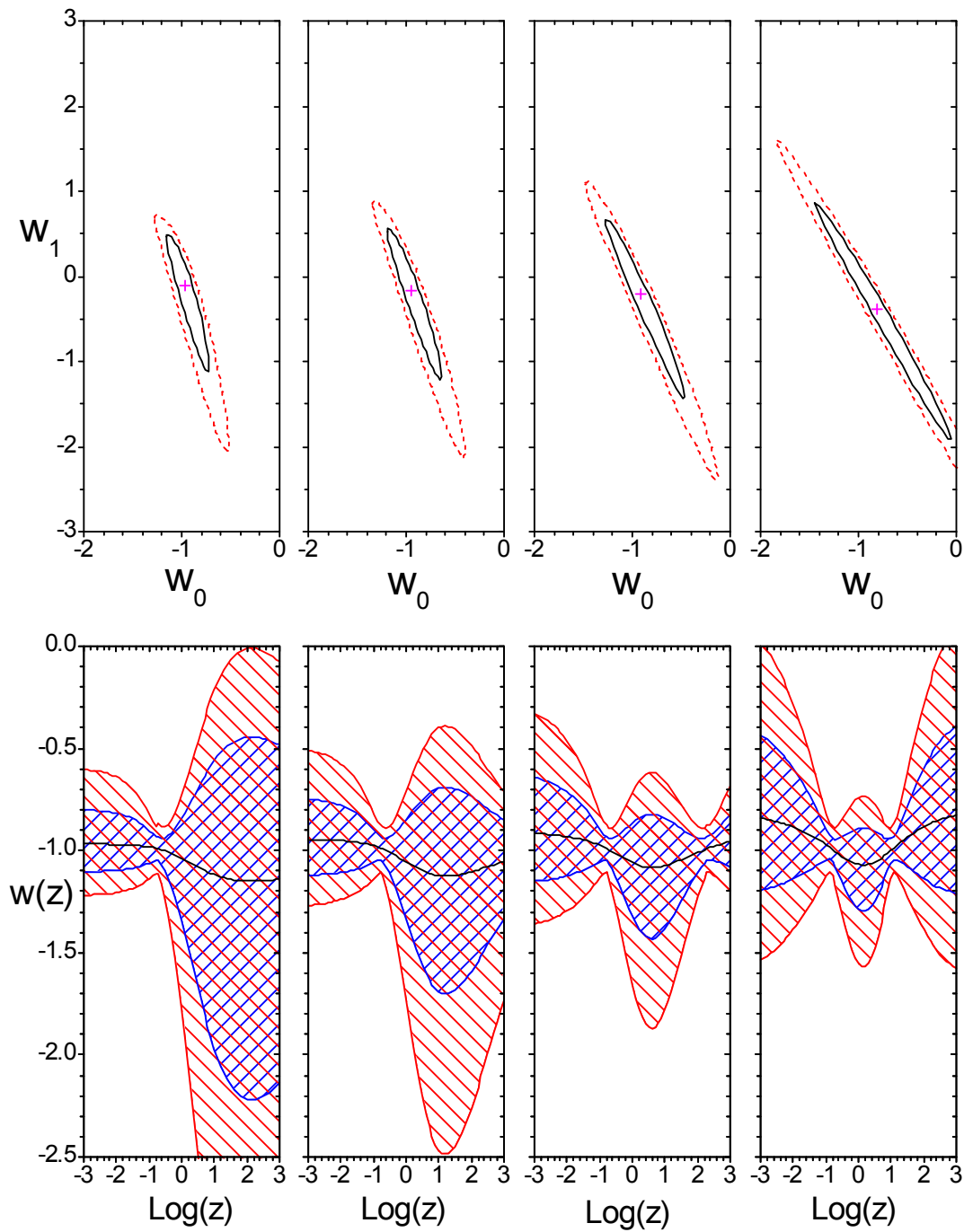}
\figcaption{The $68\%$ and $95\%$ confidence contours of the logarithmic form model
$w(z) = w_0  + w_1 {\rm ln}(1+z)/z^{\alpha}$ with ${\alpha}=1/5$, $1/3$,$1/2$ and $2/3$, from left to right.
The solid lines in bottoms panels show the best fit results.
The crosshairs in top panels mark the best-fit points ($w_0, w_1)= (-0.97, -0.10)$, $(-0.94, -0.16)$,
$(-0.91, -0.22)$ and $(-0.80, -0.38)$,  from left to right. \label{fig:log2}}
 \end{figure}

In figure \ref{fig:log2}, we present the allowed regions for the model IIc with the values of extra parameter
 ${\alpha}=1/5$, $1/3$, $1/2$ and $2/3$. The crosshairs in top panels mark the best-fit points
($w_0, w_1)= (-0.97, -0.10)$, $(-0.94, -0.16)$, $(-0.91, -0.22)$ and $(-0.80, -0.38)$, from left to right, respectively.
The solid lines in bottom panels show the best-fit results of $w(z)$. It is different from the models IIa and IIb but similar
to the type I models that the two parameters $w_0$ and $w_1$ have negative correlations.
It can been seen that for a larger value of ${\alpha}$, the pattern of allowed
region changes remarkably, the allowed region between two narrow nodes becomes smaller and
the allowed regions on both sides of redshift are enlarged.

\begin{figure}
\epsscale{0.9} \plotone{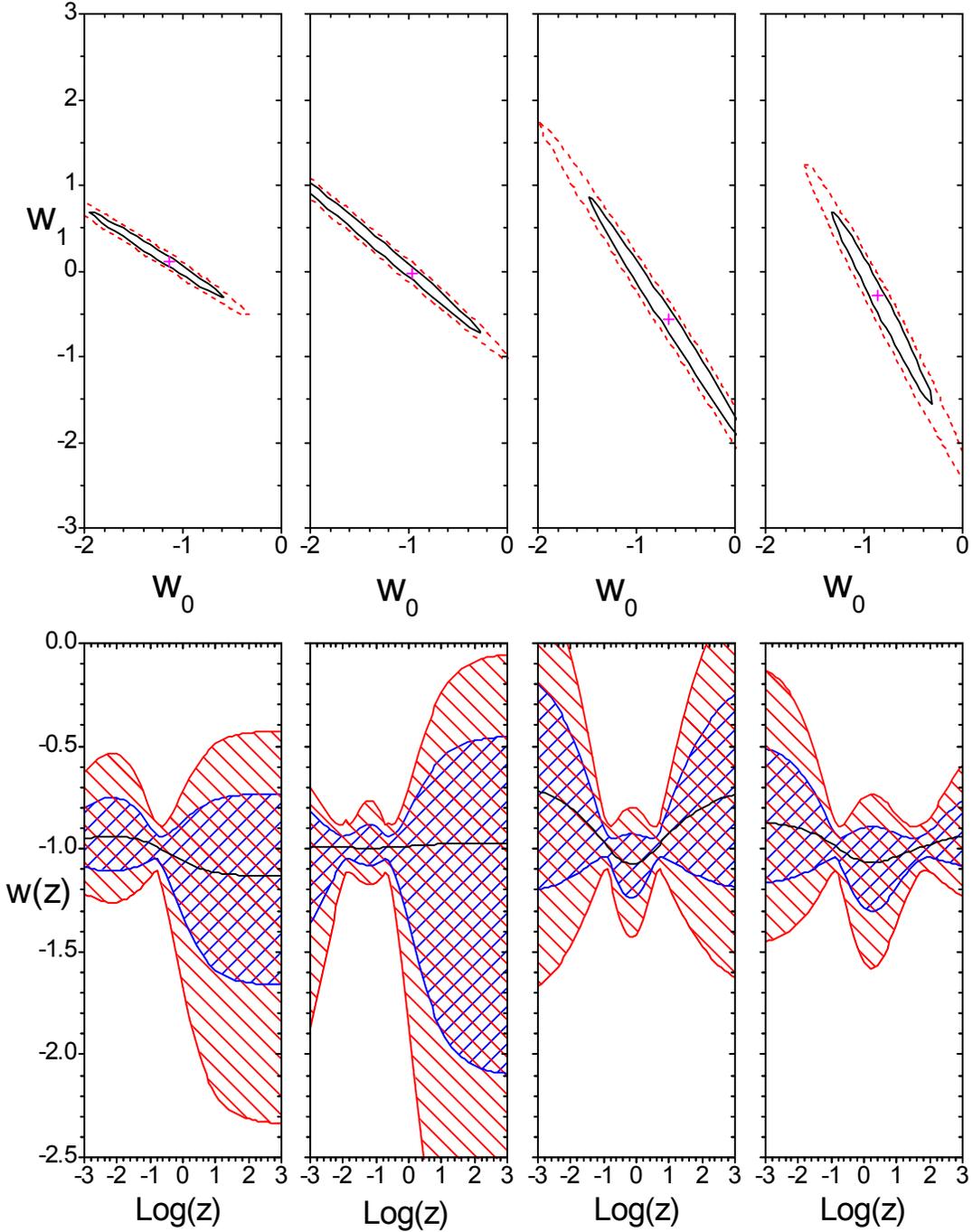}
\figcaption{The $68\%$ and $95\%$ confidence contours of the logarithmic form model
$w(z) = w_0  + w_1 z^{\alpha} {\rm ln}[(1+z)/z]$ with ${\alpha}=1/5$, $1/3$,$2/3$ and $4/5$, from left to right.
The solid lines in bottoms panels show the best fit results.
The crosshairs in top panels mark the best-fit points ($w_0, w_1)=(-1.13, -0.10)$, $(-0.97, -0.02)$,
$(-0.68, -0.56)$ and $(-0.87, -0.28)$, from left to right.  \label{fig:log3}}
 \end{figure}

The constraint results of the model IId are shown in figure \ref{fig:log3}
with values ${\alpha}=1/5$, $1/3$, $2/3$  and $4/5$. The crosshairs in top panels mark the
best-fit points ($w_0, w_1)=(-1.13, -0.10)$, $(-0.97, -0.02)$, $(-0.68, -0.56)$ and $(-0.87, -0.28)$,
from left to right. The solid lines in bottom panels show the best-fit results of $w(z)$.
It is seen that the two parameters $w_0$ and $w_1$ have negative correlations.
For different values of ${\alpha}$, the pattern of allowed region changes rapidly  and the best-fit values
of $(w_0, w_1)$ have an extremum point in the range $0<\alpha <1$.

\begin{figure}
\epsscale{0.9} \plotone{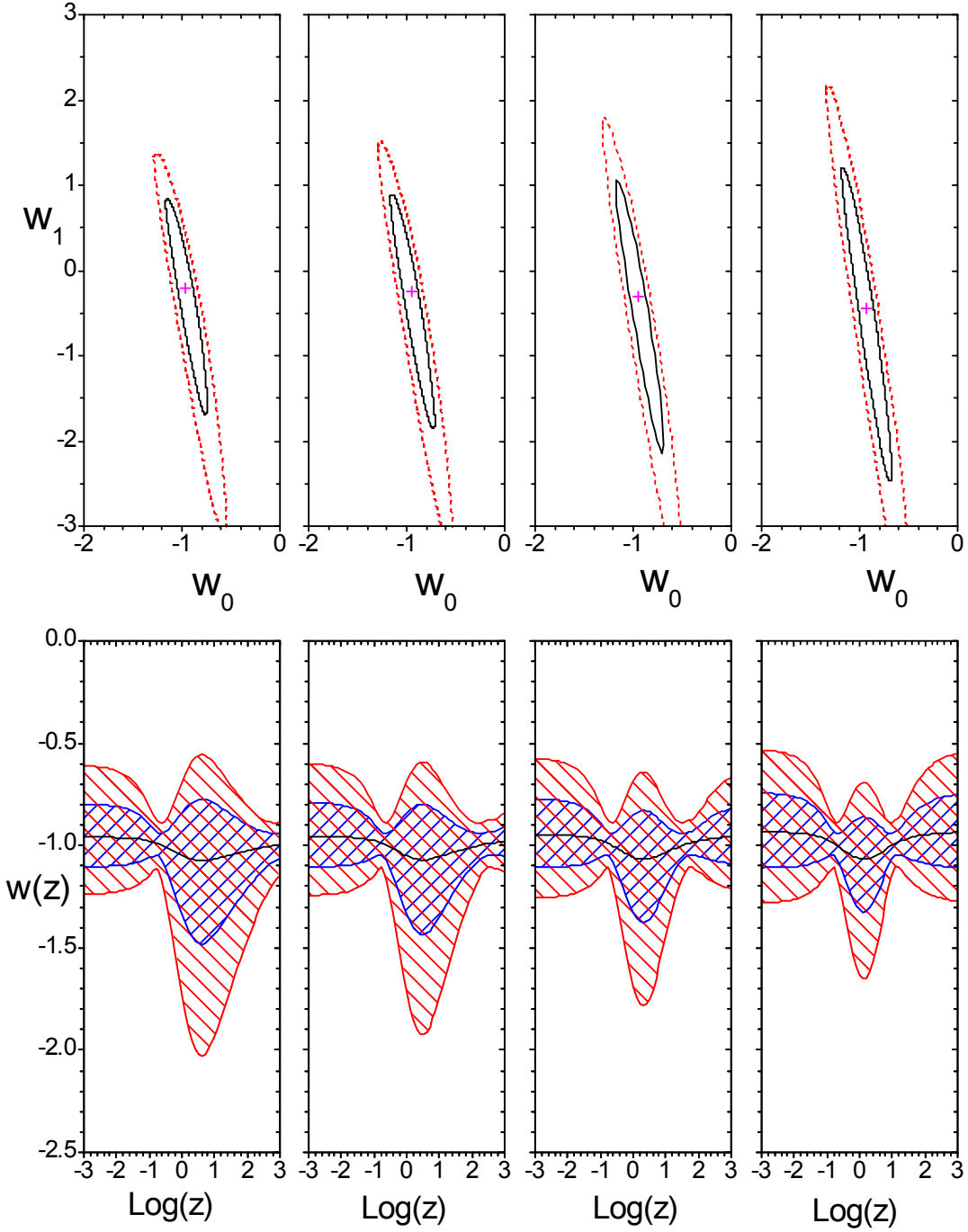}
\figcaption{The $68\%$ and $95\%$ confidence contours of the  model
$w(z) = w_0  + w_1 z/(1+z)^{\alpha} $ with ${\alpha}=5/4$, $4/3$, $3/2$, $7/4$, from left to right.
The solid lines in bottoms panels show the best fit results.
The crosshairs in top panels mark the best-fit points ($w_0, w_1)=(-0.96, -0.21)$, $(-0.95, -0.24))$, $(-0.94, -0.31)$
and $(-0.93, -0.44)$, from left to right. \label{fig:nl1}}
 \end{figure}

Figure \ref{fig:nl1} shows the constraint results of the model IIe with ${\alpha}=5/4$, $4/3$, $3/2$, $7/4$,  from left to right.
The crosshairs in top panels mark the best-fit points ($w_0, w_1)=(-0.96, -0.21)$, $(-0.95, -0.24))$, $(-0.94, -0.31)$
and $(-0.93, -0.44)$. The solid lines in bottom panels show the best-fit results.
It is seen that in all cases there are two minimum values of allowed regions and the one with small redshift
is fixed to be around $z \sim 0.2$, and the constraining power at low redshift $z<<1$ is
similar to that at high redshift $z>>1$. The allowed patterns are relatively insensitive to the change of ${\alpha}$.

\subsection{Parameter Strongly Correlative Models}

\begin{figure}
\epsscale{0.9} \plotone{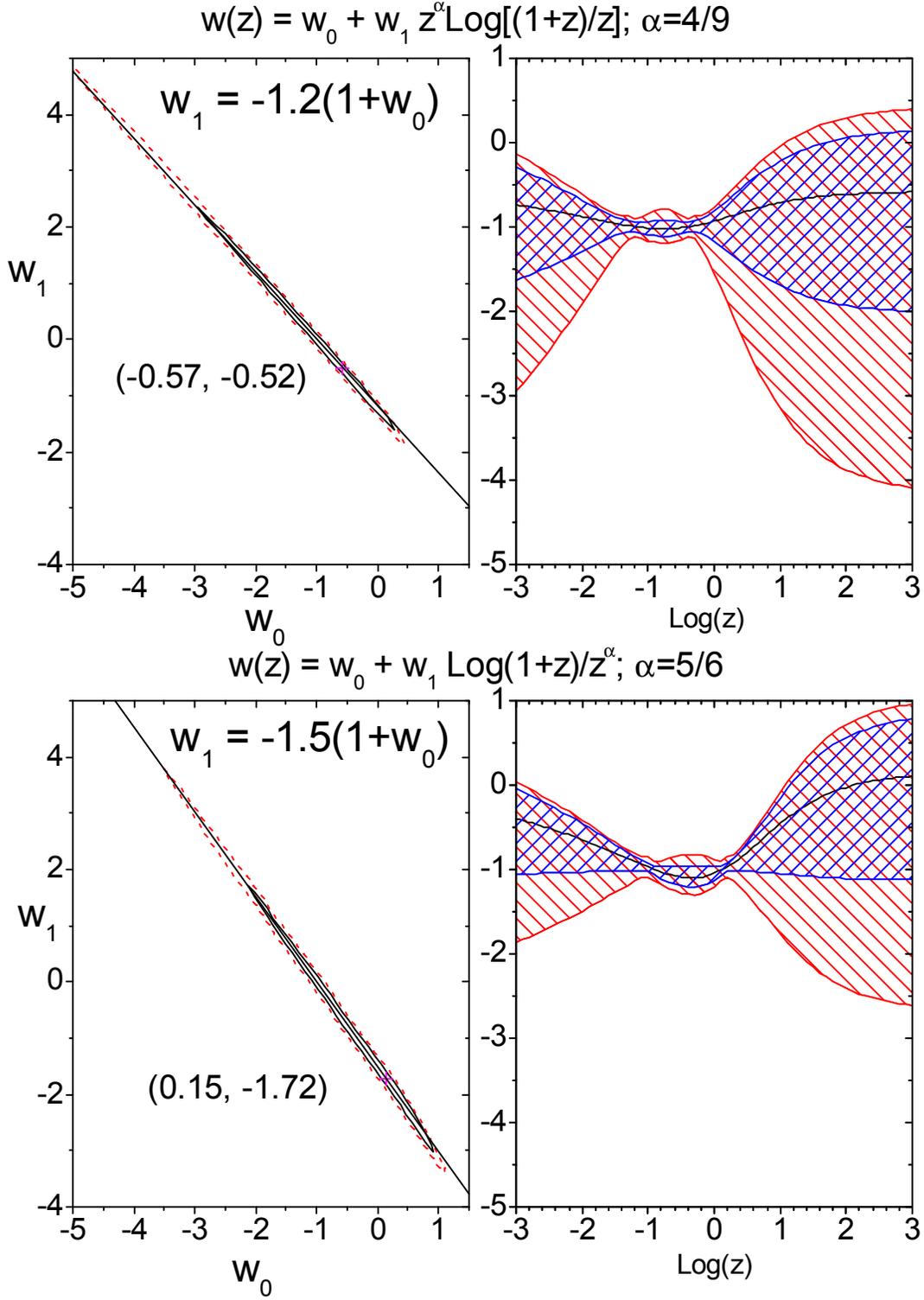}
\figcaption{The $68\%$ and $95\%$ confidence contours of the two most correlation models,
$w(z) = w_0  + w_1  z^{4/9} {\rm ln}[(1+z)/z]$ and $w(z) = w_0  + w_1  {\rm ln}(1+z)/z^{5/6}$, from top to bottom.
The solid lines in the two left panels are the linear least-squares regression line: $w_1 = -1.2 (1 + w_0)$ in top-left
panel and  $w_1 = -1.5 (1 + w_0)$ in bottom-left panel. \label{fig:cor}}
 \end{figure}

In two parameter models, it was tried to find the uncorrelated parameters through the redefinitions
 of the two initial parameters (Albrecht et al. 2006; Wang 2008). Alternatively, from the constraint results
of above two-parameter models with varying the extra parameter $\alpha$, we find two special cases with
very strong correlation between the two parameters $w_0$ and $w_1$.  It is shown in figure \ref{fig:cor}
that two models with strong parameter correlation are found from the constraint results to be
 \beq
 w(z) &=& w_0  + w_1  z^{4/9} {\rm ln}[(1+z)/z] \\
 w(z) &=& w_0  + w_1  {\rm ln}(1+z)/z^{5/6}.
 \eeq
In the left panels, we plot the $68\%$ and $95\%$ confidence contours of ($w_0, w_1$) with the crosshairs
marking the best-fit points ($w_0, w_1)=(-0.57, 0.52)$ and $(0.15, -1.72)$, from top to bottom.
It can be seen that the parameters $w_0$ and $w_1$ are strongly correlated. The linear least-squares fitting gives
the following relations
 \beq
 w_1 &=& -1.2 (1 + w_0)  \\
 w_1 &=& -1.5 (1 + w_0).
 \eeq  It is interesting to notice that at $w_0 = -1$, then $w_1=0$ and $w(z)\equiv -1$.
From the two right panels, it can be seen that in both models, the allowed regions between two narrow nodes around $z\sim1$
are quite small and the constraining powers at low and high redshift become weaker. The situation is in general expected from the
following considerations: (a) in the region with very low redshift $z<<1$, the effect of dark energy on the space-time geometry
can not be accumulated enough to show the differences among different models; (b) at high redshift $z>>1$, due to the relative
low density of dark energy based on the observations for the evolution of the early universe, it makes the data fairly insensitive
to the model variety; (c) in the region around $z\sim1$, there are more observational data and also the different
dark energy models lead  obviously to different cosmological distances. Therefore, the most sensitive redshift region of
cosmological observations on dark energy EOS $w(z)$ is around $z\sim1$, which is corresponding to the most strong constraining power.

\section{CONCLUSIONS}

We have investigated a set of time-varying dark energy models by parametering the EOS $w(z)$ via two
parameters and some of them depend on a given extra parameter. According to the boundary behavior and the local extremum
point of the EOS $w(z)$, they are classified into two types of models. From the joint analysis of four observations (SNe + BAO + CMB + $H_0$), we have presented the constraint results for all the time-varying dark energy EOS $w(z)$.

Our two-parameter models are constructed with the simple conditions that the dark energy
EOS $w(z)$ has finite boundary values at $z=0$ and $z=\infty$, and the $w(z)$ function has no more than one
local extremum point in the redshift range $0<z<\infty$. The low redshift ($z\rightarrow 0$)
 and high redshift ($z \rightarrow \infty$) behaviors of dark energy EOS $w(z)$
 are described by two parameters, and the local extremum property and boundary behavior
together characterize the pattern of $w(z)$ for its allowed region. For different given values of an extra
parameter ${\alpha}$, we have carefully studied the $w(z)$ evolution and found out its reliable values.

Based on the joint analysis of (SNe + BAO + CMB + $H_0$), we have shown for all models the best-fit results of five parameters
($w_0,w_1, \Omega_M,h,100\Omega_bh^2$) and the corresponding minimum $\chi^2_{min}$, as well as the
constraint results of $(w_0, w_1)$ with other three cosmological parameters marginalized.
It have been seen that all $\chi^2_{min}$ are close to $\chi^2_{min}=469$, which indicates that the present observational data
are not sufficient to distinguish the two-parameters models from the $\chi^2_{min}$ analysis. For all models, the
constraint results for three cosmological parameters ($\Omega_M,h,100\Omega_bh^2$)
are close to (0.28, 0.70, 2.24), though the best-fit results of $w_0$ and$w_1$ are quite different.

The three type I models have been shown to have similar functional behaviors with the CPL model and the constraint
results show their common features: (a) the two parameters $w_0$ and $w_1$ have negative correlation;
(b) there is a narrow node at $z\sim 0.2$ in the allowed regions; (c) the constraining power
at high redshift $z\rightarrow \infty$ is somewhat weaker than that at low redshift $z\sim1$;
(d) the best-fit $w(z)$ is quite flat and close to $-1$. It has been seen that our model Ia gets a stronger constraint
than the CPL model, thus the Model Ia may provide an interesting dark energy model from the best constraint results.

For the type II dark energy models which have a local extremum point at $z>0$,
we have shown the following interesting features for four of the models with the logarithmic function form:
(a) the confidence intervals of $w_1$ is much bigger than that
of $w_0$, which indicates that the observational data are less sensitive to $w_1$; (b) the allowed regions have two minimum points in the most of cases; (c) for the model IIa, the fitted $w(z)$ patterns are not sensitive to the change of the extra parameter ${\alpha}$; (d) when changing the extra parameter ${\alpha}$, the variety of the allowed region between two narrow nodes is opposite to the regions on both sides. For the model IIe without involving logarithmic function form,  it has been found that the fitted $w(z)$ patterns are relatively insensitive to the extra parameter ${\alpha}$.

From the joint analysis, we have found two interesting models which correspond to two special values of the extra parameter $\alpha$ in models IIc and IId, where the two parameters $w_0$ and $w_1$ get strong correlation with almost a linear relation: $w_1 \propto (1+w_0)$. It is also seen that the allowed regions of $w(z)$ between two narrow nodes are quite small and the constraining powers at low and high redshift become weak. The situation is understood from the fact that: (a) in the very low redshift $z<<1$, the effect of dark energy on the space-time geometry
cannot be accumulated enough to show the differences among different models; (b) at high redshift $z>>1$, the quite
low density of dark energy, which is required by the observed evolution of the early universe, makes the data fairly insensitive
to the model variety; (c) the most observational data are available in the region around $z\sim 1$, where different
dark energy models can obviously lead to different cosmological distances.

In conclusion, we have introduced a set of two-parameter dark energy models for the EOS $w(z)$ and made
a joint analysis based on the four observations (SNe + BAO + BAO + $H_0$). We have shown that the constraint results for all models get almost the same $\chi^2_{min}$ and the cosmological parameters $(\Omega_M, h, \Omega_bh^2)$, although the allowed regions for the EOS $w(z)$ are sensitive to different models and a given extra model parameter. In particular, we have found two models in which two parameters $w_0$ and $w_1$ get strong correlation and are given almost by a linear relation $w_1 \propto (1+w_0)$.

\section*{Acknowledgments}

This work is supported in part by the National Basic Research Program
of China (973 Program) under Grants No. 2010CB833000; the National
Nature Science Foundation of China (NSFC) under Grants No. 10975170,
No. 10821504 and No. 10905084; and the Project of Knowledge Innovation
Program (PKIP) of the Chinese Academy of Science.

\clearpage


\clearpage



\begin{thebibliography}{}
\bibitem[Albrecht et al., 2006)]{Albrecht2006}
Albrecht, A., Bernstein, G., Cahn, R., Freedman, W.L., Hewitt, J., Hu, W., Huth, J.,  Kamionkowski, M.,
Kolb, E.W., Knox, L., Mather, J.C., Staggs, S. \& Suntzeff, N.B., Report of the Dark Energy Task Force, astro-ph/0609591

\bibitem[Bassett, Corasaniti and Kunz (2004)]{BCK2004}
Bassett, B. A., Corasaniti, P. S. \& Kunz, M., ApJ. 617, 1 (2004)

\bibitem[Bassett et al., 2003]{bassett2003}
Bassett, B. A.,  Kunz, M., Silk, J. \&  Ungarelli, C., Phys.
Rev. D68, 043504 (2003).

\bibitem[Chevallier and Polarski (2001)]{CP2001}
Chevallier, M. \& Polarski, D., int. J. Mod. Phys. D10, 213 (2001)

\bibitem[Cole et al., 2005]{Cole2005}
Cole S., et al., MNRAS, 362, 505 (2005)

\bibitem[Colless et al., 2003]{Colless2003}
Colless M., et al., 2003, astro-ph/0306581

\bibitem[Copeland et al., 2006]{copeland2006}
Copeland, E.J., Sami, M., \&  Tsujikawa, S., Int.J.Mod.Phys.D15,1753(2006)

\bibitem[Corasaniti  and Copeland, 2003]{corasaniti2003}
Corasaniti, P. S. \& Copeland, E. J., Phys. Rev. D67, 063521  (2003)

\bibitem[Corasaniti et al., 2004]{corasaniti2004}
Corasaniti, P. S., Kunz, M., Parkinson, D., Copeland E.J.,
\& Bassett, B.A., Phys. Rev. D70, 083006 (2004).


\bibitem[Davis et al., 2007]{davis2007}
Davis, T. M. et al., ApJ, 666, 716 (2007)

\bibitem[Efstathiou, 2000]{Efstathiou2000}
Efstathiou, G., MNRAS, 342, 810 (2000)


\bibitem[Eisenstein et al., 2005]{eisenstein2005}
Eisenstein, D. J. et al., ApJ, 633, 560 (2005)

\bibitem[Eisenstein and Hu (1998)]{Eisenstein1998}
Eisenstein D. J., \& Hu W., ApJ, 496, 605 (1998)

\bibitem[Gerke and Efstathiou, 2002]{Gerke2002}
Gerke, B. F. \& Efstathiou, G., MNRAS, 335, 33 (2002)


\bibitem[Greenhill et al., 2009]{Greenhill2009}
Greenhill, L. J., Humphreys, E. M. L., Hu, W., Macri, L., Masters, K., Hagiwara, Y., Kobayashi,
H., Murata, Y., 2009, arxiv:0902.4255

\bibitem[Guy et al. (2005)]{Guy2005}
Guy, J., Astier, P., Nobili, S., Regnault, N., \& Pain, R., A\&A, 443, 781 (2005)

\bibitem[Herrnstein et al., 1999]{Herrnstein1999}
Herrnstein, J. R., et al., Nature, 400, 539 (1999)

\bibitem[Hicken et al., 2009]{Hichen2009}
Hicken et al., ApJ, 700, 1097 (2009)

\bibitem[Huey et al., 1999]{huey1999}
Huey, G.,  Wang, L.-M., Dave, R.,  Caldwell, R. R.  \& Steinhardt, P. J., Phys. Rev. D 59, 063005 (1999)

\bibitem[Humphreys et al., 2008]{Humphreys2008}
Humphreys, E. M. L., Reid, M. J., Greenhill, L. J., Moran, J. M., \& Argon, A. L., ApJ, 672,
800 (2008)

\bibitem[Huterer and Turner, 2001]{huterer2001}
Huterer, D. \& Turner, M. S.,  Phys. Rev. D64, 123527 (2001)



\bibitem[Jassal, Bagla and Padmanabhan (2005)]{Jassal2005}
Jassal, H. K., Bagla, J. S., \& Padmanabhan, T., MNRAS, 356, 11 (2005)

\bibitem[Kessler et al., 2009]{Kessler2009}
Kessler, R. et al., ApJ. 185, 32 (2009)

\bibitem[Komatsu et al., 2009]{komatsu}
Komatsu, E.  et al., ApJ. Suppl. 180, 330 (2009)

\bibitem[Komatsu et al., 2010]{wmpa7}
Komatsu, E. et al., arXiv:1001.4538

\bibitem[Kowalski et al., 2008]{Kowalski2008}
Kowalski, M., et al., ApJ, 686, 749 (2008)


\bibitem[Linder (2003)]{linder2003}
Linder, E. V.,  Phys. Rev. Lett. 90, 91301 (2003)

\bibitem[Linder and Huterer, 2005]{linder2005}
Linder, E.V.,  \& Huterer, D., Phys. Rev. D72. 043509 (2005)

\bibitem[Macri et al., 2006]{Macri2006}
Macri, L. M., Stanek, K. Z., Bersier, D., Greenhill, L. J., \& Reid, M. J., ApJ, 652, 1133 (2006)

\bibitem[Nesseris and  Perivolaropoulos (2005)]{Nesseris2005}
Nesseris, S. \& Perivolaropoulos, L.,  Phys. Rev.D 72, 123519 (2005)

\bibitem[Percival et al. (2010)]{Percival2010}
Percival, W. J.  et al., MNRAS, 401, 2148 (2010)


\bibitem[Perivolaropoulos (2005)]{Perivolaropoulos2005}
Perivolaropoulos, L., Phys.Rev. D 71, 063503 (2005)

\bibitem[Pogosian et al. (2005)]{pogosian2005}
Pogosian et al. Phys.Rev. D72, 103519(2005)


\bibitem[Riess et al., 2004]{riess2004}
Riess, A.G., et al. ApJ. 607, 665  (2004)

\bibitem[Riess et al. (2009)]{riess2009}
Riess, A. G., et al., ApJ. 699, 539 (2009)

\bibitem[Stern et al. (2010)]{Stern2010}
Stern, D., et al., JCAP. 1002, 008 (2010)

\bibitem[Wang (2008)]{wang2008}
Wang, Y., Phys. Rev. D 77, 123525 (2008)

\bibitem[Wang et al., 2000]{wanglm2000}
Wang, L.-M., Caldwell, R. R., Ostriker, J.P.  \& Steinhardt, P. J., ApJ, 530, 17 (2000)


\bibitem[Weller and Albrechet, 2002]{weller2002}
Weller, J \& Albrecht, A, Phys. Rev. D65, 103512 (2002)

\bibitem[Wood-Vasey et al., 2007]{Wood-Vasey2007}
Wood-Vasey, W. M. et al.,, ApJ, 666, 694w (2007)


\bibitem[York et al., 2000]{York2000}
York, D.G., et al., ApJ, 120, 1579 (2000)

\bibitem[Zhang et al., 2009]{zhang2009}
Zhang, Q.J., Cheng, L.M. \& Wu, Y.L.,  2009, ApJ, 694, 1402

\bibitem[Zhang and Wu, 2010]{zhang2010}
Zhang, Q.J. \& Wu, Y.L.,  2010, JCAP, 08, 038


\end{thebibliography}
\end{document}